\documentclass[12pt,a4paper]{article}
\usepackage{subfigure}
\usepackage{subfigure}
\usepackage{epsfig}
\usepackage{times}
\title{\bf The Galactic O-Star Spectral Survey (GOSSS) \\ Project status and first results
}
\author{Alfredo Sota$^1$, Jes\'us Ma\'{\i}z Apell\'aniz$^1$, Rodolfo H. Barb\'a$^2$, \\ Nolan R. Walborn$^3$, Emilio J. Alfaro$^1$, Roberto C. Gamen$^4$, \\ Nidia I. Morrell$^5$, Julia I. Arias$^2$ and Miguel Penad\'es Ordaz$^1$\\
\vspace{1cm}\\
\normalsize $^1$ Instituto de Astrof\'{\i}sica de Andaluc\'{\i}a-CSIC, Glorieta de la Astronom\'{\i}a s/n, 18008 Granada, Spain\\ 
\normalsize $^2$ Departamento de F\'{\i}sica, Universidad de La Serena, Benavente 980, La Serena, Chile\\
\normalsize $^3$ Space Telescope Science Institute, 3700 San Martin Drive, Baltimore, MD 21218, USA\\
\normalsize $^4$ Instituto de Astrof\'{\i}sica de La Plata-CONICET, Paseo del Bosque s/n, 1900 La Plata, Argentina\\
\normalsize $^5$ Las Campanas Observatory, Observatories of the Carnegie Institution of Washington, La Serena, Chile}
\date{\mbox{}}
\begin{document}
\maketitle
\pagestyle{empty}
%
%
\def\bull{\vrule height .9ex width .8ex depth -.1ex}
\makeatletter
\def\ps@plain{\let\@mkboth\gobbletwo
\def\@oddhead{}\def\@oddfoot{\hfil\tiny\bull\quad
``The multi-wavelength view of hot, massive stars''; 39$^{\rm th}$ Li\`ege Int.\ Astroph.\ Coll., 12-16 July 2010 \quad\bull}%
\def\@evenhead{}\let\@evenfoot\@oddfoot}
\makeatother
%
%
\def\beginrefer{\section*{References}%
\begin{quotation}\mbox{}\par}
\def\refer#1\par{{\setlength{\parindent}{-\leftmargin}\indent#1\par}}
\def\endrefer{\end{quotation}}
%
%
{\noindent\small{\bf Abstract:} 
The Galactic O-Star Spectroscopic Survey (GOSSS) is a project that is observing all known 
Galactic O stars with $B$ $<$ 13 ($\sim$2000 objects) in the blue-violet part of the spectrum 
with R$\sim$2500. It also includes two companion surveys (a spectroscopic one at 
R$\sim$1500 and a high resolution imaging one). It is based on v2.0 of the  
Galactic O star catalog (v1, Ma\'{\i}z Apell\'aniz et al. \ 2004; 
v2, Sota et al. \ 2008). We have completed the first part of the main project. Here we present 
results on the first 400 objects of the sample.
}
%
%
\section{Description}
The Northern part of the survey is being carried out from the Sierra Nevada and Calar Alto 
observatories (Spain) and the Southern part from Las Campanas (Chile). Although the data 
have been acquired at three different observatories, they have nearly identical characteristics: 
Uniform and high signal to noise ratio (200-300), same spectral resolution (R $\sim$2500), and 
similar spectral coverage ($\sim$3900 to 5100 \AA). To date, we have completed the first part of 
the project (observing the first 400 objects of the sample). The main objective is to publish a 
new Galactic O-Star Atlas as well as the spectrograms for all stars. Figures \ 1 and \ 2 show two atlas sheet with 
the same  luminosity class and the same spectral type. We have 
also observed more than 300 objects from the second part of the project, a number that is 
increasing in current campaigns. For each star, we typically have two or more epochs. The survey 
will be used for a number of purposes, such as a precise determination of the IMF for massive 
stars, the measurement of radial velocities for Galactic kinematical studies, and the detection of 
unknown massive binaries. Results will be made available through a dedicated web server, will be 
incorporated into the Virtual Observatory, and will include the most complete spectral library of 
massive stars to date.
A paper with the first $\sim$180 northern stars of the survey, including a new spectral classification atlas, will be presented by Sota et al. (2010,  submitted to ApJS). Another paper with $\sim$200 southern stars will follow next year.

\begin{figure*}[!h]
\centering
\includegraphics[width=11.2cm]{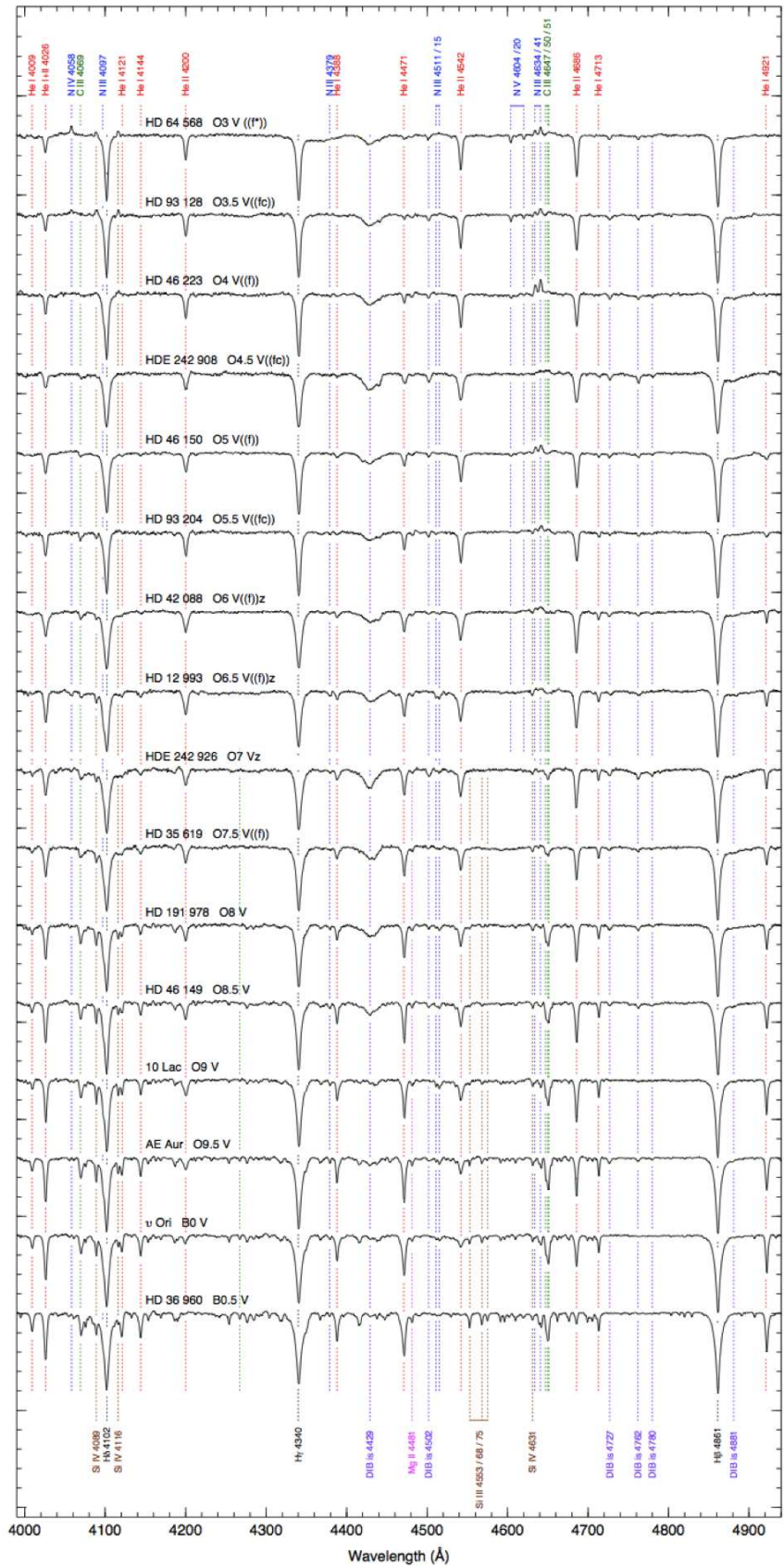}
\caption{Spectrograms for luminosity class V of Galactic O stars. \label{fig_1}}
\end{figure*}

\begin{figure*}[!h]
\centering
\includegraphics[width=13cm]{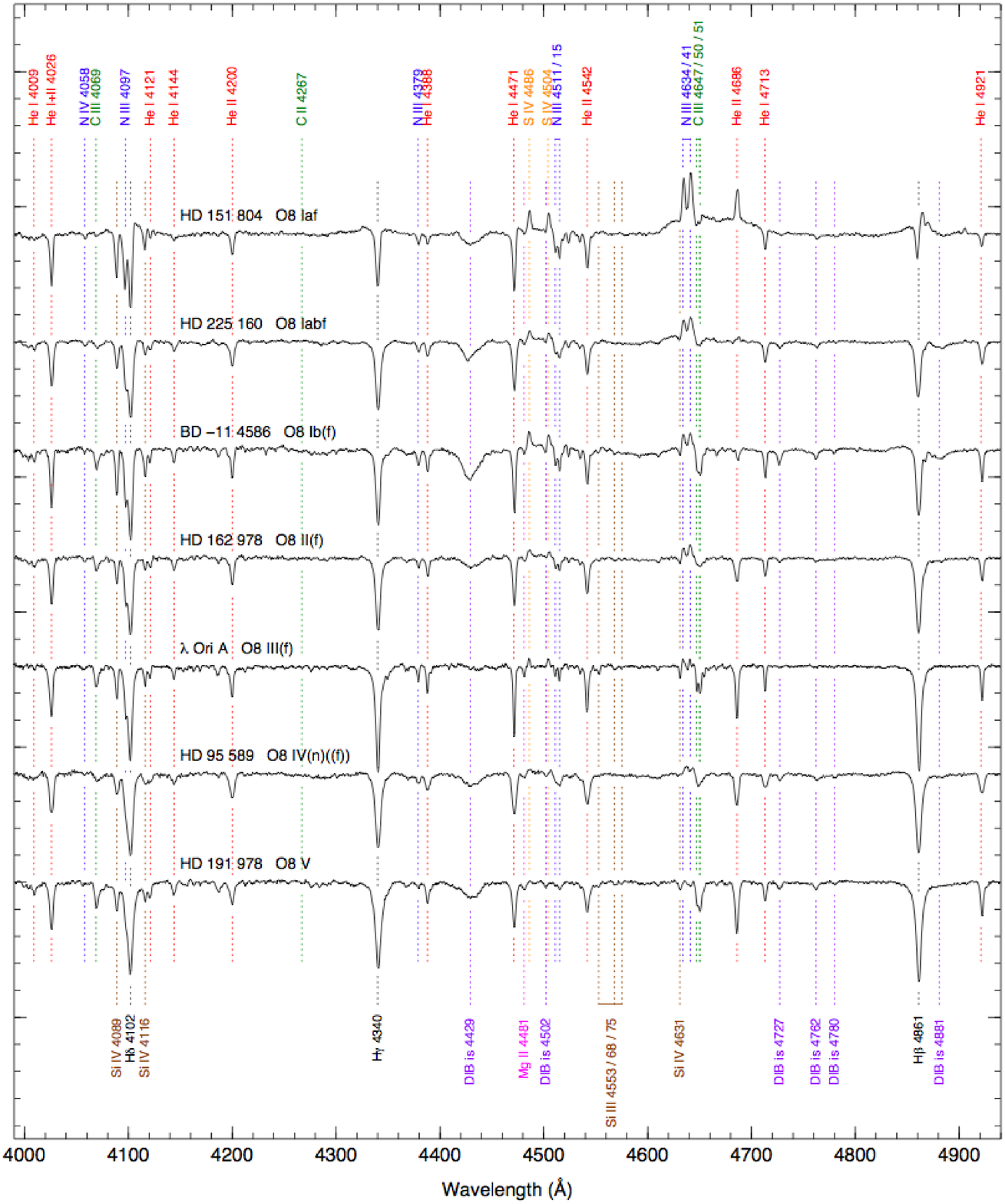}
\caption{Luminosity effects at spectral type O8. \label{fig_2}}
\end{figure*}

\section{C\,{\sc iii} Emission Lines in Of Spectra}
We introduce the Ofc category, which consists of normal spectra with 
C\,{\sc iii} $\lambda$$\lambda$4647-4650-4652 emission lines of comparable intensity to those of 
the Of defining lines N\,{\sc iii} $\lambda$$\lambda$4634-4640-4642 (Figure 3b). The former feature 
is strongly peaked to spectral type O5, at all luminosity classes, but preferentially in some associations 
or clusters and not others. This behavior contrasts with that of the selective C\,{\sc iii} $\lambda$5696 
emission, which has a much wider spectral-type distribution. It is also distinct from that of the Of?p stars, 
which have C\,{\sc iii} $\lambda$$\lambda$4647-4650-4652 emission (localized to some particular region of 
the unknown circumstellar structures), but otherwise peculiar and variable spectra. Magnetic fields have 
recently been detected on three members of the latter category. We present two new extreme Of?p 
stars, NGC 1624-2 and CPD -28$^{\circ}$ 2561, bringing the number known in the Galaxy to five (Figure 3a). 
Modeling of the behavior of these spectral features can be expected to better define the physical parameters of 
both normal and peculiar objects, as well as the atomic physics involved (see Walborn, et al. \ 2010).

\begin{figure}[!h]
 Ê Ê \centering
 Ê Ê \subfigure[]
 Ê Ê Ê Ê {\label{fig4}}
 Ê Ê Ê Ê Ê\includegraphics[width=.4\textwidth, angle= 0]{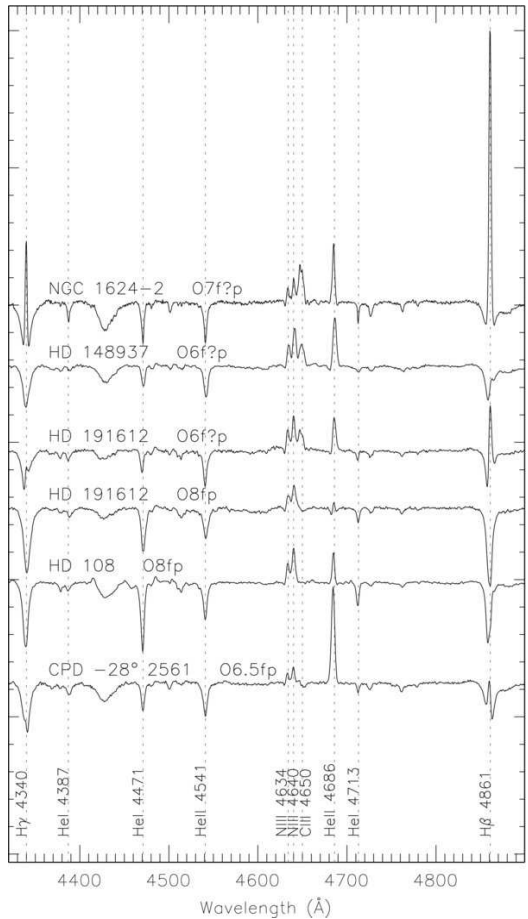}
 Ê Ê Ê Ê Ê\hspace{-0.1in}
 Ê Ê \subfigure[]
 Ê Ê Ê Ê Ê{\label{fig5}}
 Ê Ê Ê Ê Ê\includegraphics[width=.4\textwidth, angle = 0]{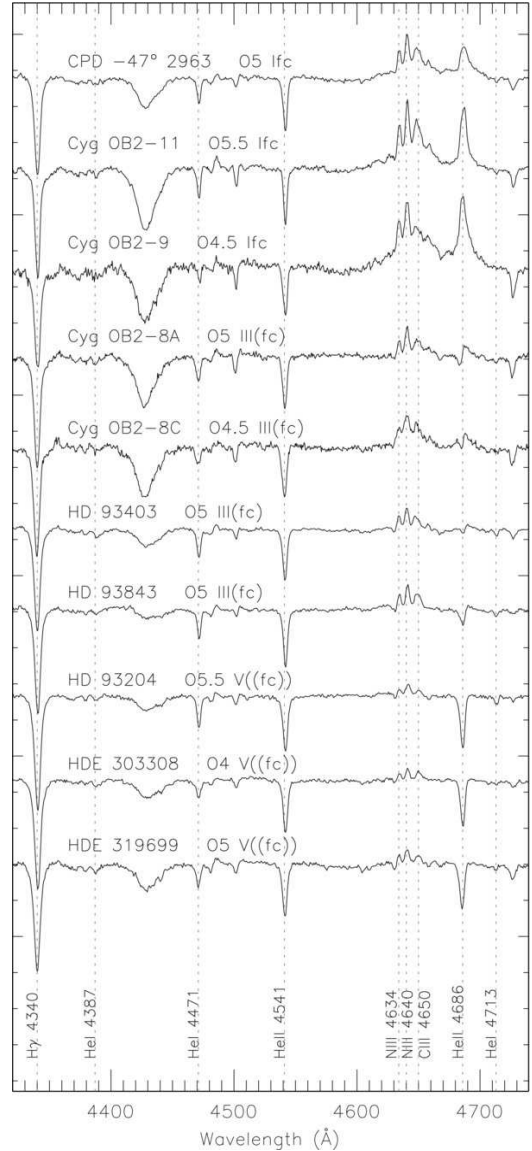}
 Ê Ê \caption{a)  Spectra of the Galactic Of?p stars in the blue-green region, and b)  Luminosity sequence of Ofc spectra in the blue-green region. }
 Ê Ê \label{fig4y5}
\end{figure}



\section{Data Reduction}
Due to the large amount of data that we are collecting, we have developed a pipeline for the automatic 
(quicklook) or semi-automatic (full) data reduction. In quicklook mode, we are able to get the rectified 
and calibrated spectrograms for all the stars that we have observed in a single night just in a few minutes 
after the observations. The pipeline can identify every kind of file and apply the standard reduction to the 
images (bias subtraction, flat-fielding, bad pixel mask, ...). It also extracts the spectrograms from the 
images (including close visual binaries; see Figure \ 4), aligns and combines all the spectrogram from the same star (to 
increase the signal to noise ratio and correct defects like cosmic rays), calibrates in wavelength and rectifies 
to the continuum. The same operations are performed in full mode with the user being able to tweak the files 
and parameters used.

\begin{figure*}[!h]
\centering
\includegraphics[width=15cm]{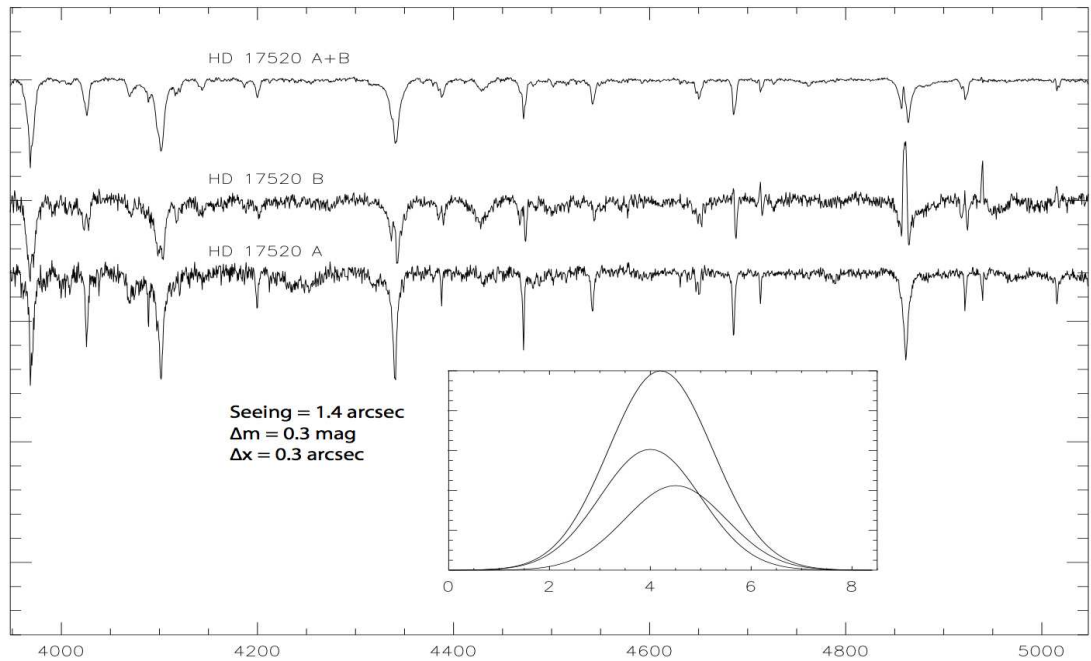}
\caption{ Example of disentangling. The visual binary HD 17520 A+B is separated by only 0$.\!\!^{\prime\prime}$3, a fraction of the seeing in this exposure. This is the most extreme case in our sample. We are able to disentangle the two pair components to show an O type spectra and a Be type spectra (pay attention to the emission lines in the Be star).  \label{fig_5}}
\end{figure*}

%
%
%
\section*{Acknowledgements}
Support for this work was provided by [a] the Spanish Government Ministerio de Ciencia e Innovaci\'on through
grants AYA2007-64052, the Ram\'on y Cajal Fellowship program, and FEDER funds; [b] the Junta de Andaluc\'{\i}a
grant P08-TIC-4075; and [c] NASA through grants GO-10205, GO-10602, and GO-10898 from the Space Telescope
Science Institute, which is operated by the Association of Universities for Research in Astronomy Inc., under
NASA contract NAS~5-26555.
%
%
\footnotesize
\beginrefer

\refer Ma\'{\i}z Apell\'aniz, J., Walborn, N. R., Galu\'e, H. \'A., \& Wei, Lisa H., 2004, ApJS, 151, 103

\refer Sota, A., Ma\'{\i}z Apell\'aniz, J., Walborn, N. R., \& Shida, R. Y., 2008, RMxAC, 33, 56

\refer Walborn, N. R., Sota, A., Ma\'{\i}z Apell\'aniz, J., Alfaro, E. J., Morrell, N. I., Barb\'a, R. H., Arias, J. I., \& Gamen, R. C., 2010, ApJ, 711L, 143

\endrefer           
\end{document}